# Clarifying tipping terminology

Niklas Boers[1,2,3*#], Robert E. Kopp[4*#], Peter Ashwin[3], Sebastian Bathiany[1,2], Maya Ben-Yami[1,2], Reyk Börner[5], Jofre Carnicer[6], Henk Dijkstra[5], Thomas L. Frölicher[7,8], Michael Ghil[9,10,11], Elisabeth Gilmore[12], Gabriele C.Hegerl[13], Laura Jackson[14], Shipra Jain[15], David Lapola[16], Timothy M. Lenton[17], Valerio Lucarini[18,19], Teng Liu[1,2], Virna L. Meccia[20], Andreas Morr[21], Anastasia Romanou[22,23], Thomas F. Stocker[7,8], Ricarda Winkelmann[2,24], Cunde Xiao[25], Shuting Yang[26]

[1]Munich Climate Center and Earth System Modelling Group, Department of Aerospace and Geodesy, TUM School of Engineering and Design, Technical University of Munich, Munich, Germany (n.boers@tum.de)
[2]Potsdam Institute for Climate Impact Research, Potsdam, Germany (boers@pik-potsdam.de)
[3]Department of Mathematics, University of Exeter, Exeter, UK
[4]Department of Earth & Planetary Sciences and Rutgers Climate & Energy Institute, Rutgers University, New Brunswick, NJ, USA
[5]Institute for Marine and Atmospheric research Utrecht, Department of Physics, Utrecht University, Utrecht, The Netherlands
[6]University of Barcelona, CREAF-IRBio, Spain
[7]Climate and Environmental Physics, Physics Institute, University of Bern, Switzerland
[8]Oeschger Centre for Climate Change Research, University of Bern, Switzerland
[9]Department of Geosciences, Ecole Normale Supérieure, Paris, France
[10]Department of Atmospheric and Oceanic Sciences, UCLA, Los Angeles, CA, USA
[11]Department of Mathematics, Imperial College London, London, UK
[12]Norman Paterson School of International Affairs and Civil and Environmental Engineering, Carleton University, Ottawa, Canada
[13]School of GeoSciences, University of Edinburgh, UK
[14]Met Office Hadley Center, Exeter, UK
[15]Department of Risk and Disaster Reduction, Faculty of Mathematical and Physical Sciences, University College London, United Kingdom
[16]University of Campinas - UNICAMP, Campinas SP 13083-886, Brazil
[17]Global Systems Institute, University of Exeter, Exeter, UK
[18]School of Computing and Mathematical Sciences, University of Leicester, Leicester, UK
[19]School of Sciences, Great Bay University, Dongguan, P.R. China
[20]National Research Council of Italy, Institute for Atmospheric Sciences and Climate, Bologna, Italy.
[21]Department of Mathematics, TUM School of Computation, Information and Technology, Technical University of Munich, Munich, Germany
[22]NASA Goddard Institute for Space Studies, New York, USA
[23]Department of Applied Physics and Applied Mathematics, Columbia University, New York, USA
[24]Department of Integrative Earth Systems Science, Max Planck Institute of Geoanthropology, Jena, Germany
[25]Key Laboratory of Polar Atmosphere-ice System for Weather and Climate, Institute of Atmospheric Sciences, Fudan University, Shanghai, China

[26]National Climate Research Centre, Danish Meteorological Institute, Copenhagen, Denmark

*These authors contributed equally.
#Corresponding authors: N. Boers (n.boers@tum.de) & R. E. Kopp (robert.kopp@rutgers.edu)

**Abstract**
The notion of 'tipping points' and related terminology have so far had inconsistent and sometimes confusing definitions in both scientific and popular literature. While it is difficult for the scientific community to align popular usage of the term 'tipping point' with a precise definition, technical usage of related terms would benefit from greater consistency and rigor. Toward this end, we offer a systematic set of terms to clarify formal communication around potential tipping dynamics, with focus on examples from Earth system dynamics.

# Introduction

Since being introduced into climate and ecological research in the early 2000s, the terminology related to tipping phenomena has varied over time and across disciplines and authors (e.g., Boers et al., 2022, 2025; Lenton et al., 2008; ; Macneill et al., 2026; Scheffer et al., 2001), which has led to considerable confusion (Kopp et al., 2025). For example, the term 'tipping point' has been used to refer to distinct but related aspects of 'tipping' dynamics. In some cases, it refers to the threshold associated with triggering a state transition in a system (Lenton et al., 2008; Milkoreit et al., 2018a; IPCC, 2021a); in other cases, it refers to the transition itself (Abrams et al., 2026); in still other cases (most often in less formal publications), it is used to refer to systems that can undergo such transitions. It is also often unclear if the term 'tipping point' refers to an internal threshold of a system (i.e., a regime boundary at which the system and its internal processes change behaviour), to a critical threshold in the level and/or rate of external drivers acting on a system, or to the projection of such critical thresholds into a one-dimensional aggregate metric such as global-mean surface air temperature (Armstrong McKay et al., 2022).

Further confusion has been caused by not clearly distinguishing (a) transitions that may be slow on decision-relevant human time scales but practically irreversible even if given sufficient time to equilibrate, from (b) strong nonlinear responses that may be rapid on human time scales, but fully reversible when given sufficient time to equilibrate. In addition, the notion of 'abruptness' often appears when describing tipping dynamics. It is often assumed that 'abruptness' relates to suddenness in time, but it is rarely specified if it is really meant with respect to time, or with respect to the forcing acting on the system in question. For example, the meltdown of the Greenland ice sheet would be abrupt – in the sense of "very steep" – when presenting the ice volume as a function of regional temperatures in a quasi-equilibrium setting, where these temperatures change sufficiently slowly to give the ice sheet time to find its equilibrium state with each temperature change. However, this transition would take centuries to millennia and would thus not be considered abrupt on human time scales. By contrast, a transition of a system that exhibits hysteresis could in theory be reverted on human timescales if the hysteresis is narrow. This has been suggested for the eutrophication of shallow lakes (Jeppesen et al., 2005; Scheffer et al., 1993), but has not been demonstrated for large-scale Earth system components. By further contrast, theoretically reversible nonlinear shifts might not be fully reversible on human time scales because of pronounced latency (see definition below) of the system response. Moreover, they could also turn out to be irreversible in practice because they might be coupled to another process that is irreversible. Because of the ambiguities in the usage of the terms abruptness and irreversibility (Buhr et al., 2025), below we propose a definition of tipping transitions without invoking them (see also Börner & Dijkstra, 2026). However, we acknowledge that they can be key traits that remain important for the practical impacts of tipping transitions and discuss the relationship between tipping transitions, abruptness and irreversibility in detail below.

In 20th-century technical literature, the term ‘tipping point’ was used almost exclusively in discussions of the dynamics of urban neighborhood segregation (e.g., Grodzins, 1957; Schelling, 1971). Its use was expanded and propagated by Malcolm Gladwell’s best-selling pop sociology book *The Tipping Point* (Gladwell, 2000). Having achieved widespread popular use, the term was then adopted into the climate and ecology literature, where it took on meanings previously associated with other terms such as ‘multiple equilibria’, ‘bifurcations,’ ‘abrupt change,’ ‘regime shifts’, ‘irreversibility’, and ‘critical transitions’ (Abrams et al., 2026; Kopp et al., 2016, 2025; Lenton et al., 2008; Russill, 2015; Russill & Nyssa, 2009; Scheffer, 2009). As it gained popularity in these new contexts, it was also increasingly employed to describe social phenomena relevant to climate impacts, technology and policy (eg Milkoreit et al., 2018). Given that the phrase came into the public discourse independently and before its application in the climate context, attempts to impose technical meanings on the term have met with mixed success at best, even as the formal conceptual foundations of the nonlinear phenomena it describes have been further developed (Abrams et al., 2026; Ashwin et al., 2012; Boers et al., 2022; Börner & Dijkstra, 2026; Ghil, 2001; Lenton et al., 2008; Lucarini & Chekroun, 2024; Milkoreit et al., 2018b; Van Nes et al., 2016; Wieczorek et al., 2023).

In the context of climate science, the term 'tipping point' has seen an inflationary usage since its introduction in the early 2000s and, at the same time, a creeping scope in both the scientific literature and public discourse. A straightforward interpretation of the etymology of the term ‘tipping point’ risks oversimplifying the nonlinear dynamics of complex systems, since its use suggests (generally inaccurately) a specific zero-dimensional threshold in a system’s dynamics. We outline below how the term could be made more precise, referring to the specific point in time when a transition is triggered. Nevertheless, in view of the existing ambiguity in the scientific literature and public perception, we recommend using other terminology as introduced in the following.

Without using ‘tipping’ terminology, the phenomenon that is today often referred to as bifurcation-induced tipping (Ashwin et al., 2012) was described in multiple instances already more than half a century ago, e.g., in the context of energy-balance models and the discovery of the Snowball Earth state (Budyko, 1969; Ghil, 1976; North, 1975; Sellers, 1969) and in the seminal work of Henry Stommel (Stommel, 1961) suggesting bistability in ocean circulation systems like the Atlantic Meridional Overturning Circulation (AMOC; (see also, e.g. (Dijkstra & Ghil, 2005; Ghil & Lucarini, 2020) for more recent treatments). Similarly, the phenomenon now called rate-induced tipping was discussed, without using the term, in the context of rapid greenhouse gas forcing leading to substantial AMOC weakening (Stocker & Schmittner, 1997; Van Westen et al., 2026) and of time-dependent hosing experiments (Lucarini et al., 2007). The phenomenon now described as noise-induced tipping between alternative metastable states was identified across a wide range of fields since the 1940s (Kramers, 1940) and highlighted in recent literature (Cini et al., 2024); see also (Romanou et al., 2025) for possible effects of extreme events on tipping events. In social systems, 'tipping' terminology also supplements more established terms, such as 'leverage points' and 'learning curves'. Sometimes, the term 'tipping' is used to describe what might be merely a nonlinear or abrupt change. These examples highlight the fact that the term 'tipping' is not essential for a precise description of the relevant phenomena. A range of terms is currently used in the community to describe phenomena that may or may not constitute tipping, with some studies avoiding the term altogether.

One characteristic of many tipping systems is deep uncertainty regarding key processes and how to model them (Kopp et al., 2025). Confusingly, tipping events are often described as 'high-impact, low-likelihood' events, even though their likelihood is not in fact known to be low, but is rather poorly known or even unknown, as is the case under deep uncertainty (IPCC, 2021b; Kopp et al., 2023, 2025; Lempert et al., 2024; Wood et al., 2023). Even the Intergovernmental Panel on Climate Change (IPCC) Sixth Assessment Report (AR6) contributed to this confusion by designating a high-end sea-level rise storyline with a significant contribution from ice-sheet tipping dynamics as a 'low-likelihood, high-impact' storyline, while at the same time clarifying in a footnote that the likelihood of such a storyline was poorly known rather than low (IPCC, 2021b).

It seems useful to define tipping behavior more generally as a form of change that is dynamically distinct from other forms of change (Börner & Dijkstra, 2026). In an attempt to impose order on tipping terminology, we propose the following definitions and offer accompanying clarifications:

# Defining tipping transitions

*A **dynamical regime** is a domain in state space from which the system returns to the same characteristic set of states via its internal stabilizing dynamics, and within which it therefore persists unless external effects drive it outside the domain.*

The boundary of a dynamical regime can only be crossed due to an externally forced change in internal dynamics, or the direct effect of an external perturbation on the system state. Domain boundaries arise from internal attracting dynamics and from interactions with external drivers and can evolve in time. In dynamical systems terminology, a dynamical regime is a basin of attraction maintained by stabilizing (negative) feedbacks, i.e. a collection of states from which the system will return to the same characteristic state (the attractor) in response to small perturbations. Dynamical regimes can **co-exist** in the sense that they exist for the same forcing but remain disjoint (multistability). We call regimes that co-exist for a certain forcing range **alternative regimes**.

*A **tipping transition** is a self-sustaining transition between alternative dynamical regimes of a system.*

Tipping transitions can be triggered by (1) an externally forced change in internal dynamics, which will influence the shape, structure, characteristics, and potentially the number of dynamical regimes of a system, or (2) individual external perturbations, which directly act on the system state at a given time, but without affecting the system's dynamical regimes. Note that the above definition hence excludes oscillations characterized by purely internal variability (e.g., modes of unforced climate variability) as tipping transitions. We emphasize that the above definition of tipping transition is not restricted to systems in equilibrium but rather includes transient dynamics (also see rate-induced transitions below).

*A **tipping system** is a system that is capable of undergoing tipping transitions. An **Earth system tipping element** is a tipping system that is also a large-scale component of the Earth system. **Tipping dynamics** are the dynamics of a tipping system that involve tipping transitions.*

Tipping transitions are self-sustaining in the sense that they will, once triggered, continue until the system has settled in the alternative regime, even if their initial trigger ends. Phenomena related to tipping transitions have in the literature often been discussed in the context of linear feedbacks, which can be positive, i.e. amplifying, or negative, i.e. damping. The linear feedback description can be formalised using response theory for nonequilibrium systems (Ghil & Lucarini, 2020; Lucarini & Chekroun, 2024) but is of limited use to describe or define tipping transitions in their full generality, including situations where the forcing acts on time scales comparable to the internal time scales of the system. A self-consistent, rigorous way of defining the effects of external drivers on systems with internal variability is provided by the theory of nonautonomous dynamical systems (Ashwin et al., 2026; Chekroun et al., 2011; Crisan et al., 2026).

There are three main categories of tipping transitions in dynamical systems theory (Ashwin et al., 2012).

1) *A **bifurcation-induced tipping transition** occurs when a system loses its current dynamical regime and consequently transitions to an alternative regime.*

This first category of tipping transitions is the classical setting which is relevant when variability is relatively small and forcing is changing slowly compared to the system's time scale; it is inherently an equilibrium phenomenon. Bifurcation-induced transitions can be understood in terms of feedbacks, since they occur when net stabilizing processes vanish once a critical threshold in the magnitude of external forcing is crossed. In this case the net feedback becomes destabilizing. Formally, a destabilizing feedback is a feedback that is not only positive (feedback factor $f > 0$), but approaches +1, leading to a diverging gain factor $g = 1/(1-f)$ (Peixoto & Oort, 1993). In the context of the ice-albedo feedback, for example, this would be associated with a transition into or out of a planetary-scale Snowball Earth glaciation (Budyko, 1969; Ghil & Lucarini, 2020; Sellers, 1969). Note that only some bifurcations known in the mathematical literature can lead to tipping transitions as defined above.

2) *A **rate-induced tipping transition** occurs when the system leaves its current regime because the regime itself changes so rapidly (without disappearing) in response to high rates of forcing that the stabilizing forces of the regime's attractor cannot keep the system within the regime (Ritchie et al., 2023).*

In the case of rate-induced transitions, the dynamics becomes self-sustained once the system crosses the regime boundary and begins to settle in the alternative regime. For example, several studies have demonstrated rate-induced transitions of AMOC across the hierarchy of model complexity (Stocker & Schmittner, 1997; Van Westen et al., 2026; Zhang et al., 2026); at high rates of forcing, the AMOC transition occurs at considerably smaller levels of freshwater or greenhouse gas forcing, because the stabilizing forces of the strong AMOC regime cannot keep up with the rapid shift of the regime itself.

3) *A **noise-induced tipping transition** occurs when the system crosses a regime boundary in response to external ('noisy') perturbations.*

In this case, self-sustained dynamics are triggered once the regime boundary is crossed and the attracting forces of the alternative regime begin to dominate. Importantly, the noise or perturbations in this case act directly on the system, rather than on the forcing. As a prominent example, the Dansgaard-Oeschger warming and cooling events (Dansgaard, 1987), which are the archetype of potential tipping transitions evidenced in paleoclimate proxy records (Boers et al., 2022), have been suggested to probably be noise-induced (Ditlevsen et al., 2007), with sea ice and wind-stress variability, as well as iceberg discharge as possible physical candidates for the stochastic perturbations. However, alternative dynamical mechanisms, such as self-sustained oscillations, have more recently been proposed (Peltier & Vettoretti, 2014; Vettoretti et al., 2022) and would not be considered noise-induced tipping transitions.

In reality, tipping transitions would typically be driven by a combination of these factors: for example, the typical case in multistable systems subject to noise perturbations is a combination of bifurcation- and noise-induced tipping in the sense that both the parameter change as well as the noise are necessary for the tipping to occur at a certain time. In case the drivers change on time scales comparable to the characteristic time scale of the system, rate-induced effects can also become relevant (Alkhayuon et al., 2019). Furthermore, a tipping transition can occur through the effect of external forcing on internal variability (Romanou et al., 2023; Van Westen et al., 2026).

For completeness, we note that the term *critical threshold* is often used to refer to a threshold in the level of external forcing that causes the internal dynamical regime to vanish (bifurcation-induced transition), or its boundary to be crossed (rate- and noise-induced transition). It is important to note, however, that not only the level of forcing matters, but also its rate of change, and more generally the time evolution of the forcing pathway. In several cases it has been shown that the critical forcing level reduces with increasing rate of change (Stocker & Schmittner, 1997; Zhang et al., 2026; Van Westen et al., 2026), which also implies that communicating a critical forcing level without reference to the rate of forcing can be misleading, and in the case of reporting critical equilibrium forcing levels, can even suggest a false sense of safety (Ben-Yami et al., in prep.). Even for a single critical forcing variable, this establishes the critical threshold as a critical curve in the magnitude-rate space, rather than a singular ‘point’. Moreover, the critical threshold can often involve more than one variable, making it a high-dimensional surface. For example, for the Amazon rainforests, relevant forcings that might define regional critical thresholds include land use change (and its spatial pattern), temperature, precipitation, fire frequency, and atmospheric $CO_2$ concentrations (Bochow & Boers, 2023; Flores et al., 2024), as well as the rates of change of these drivers. Although there is a strong desire to collapse critical thresholds to the single dimension of global-mean surface temperature, we emphasize that this is often a poor way of characterizing a system’s sensitivity.

The above makes clear why the term ‘tipping point’ would be misleading to describe relevant thresholds, regardless of whether they are internal ones given by regime boundaries, or external ones given by critical values in the level and rate of (often multivariate) forcing, because both are typically complex high-dimensional objects rather than points. The term **tipping point** can be more faithful to its etymology when defined as the point in time, along a particular trajectory, at which a regime boundary is crossed — in other words, to refer to the start of a ‘tipping transition’ rather than as a synonym for either the internal or external threshold, or the process of the tipping transition. However, it should be noted that even when defined as the point in time when a tipping transition is triggered, it remains an uncertain notion specific to that particular trajectory. It will in most contexts practically be difficult to identify, and when possible, often only retrospectively. Moreover, the crossing of the regime boundary and the transition event in the corresponding time series can be significantly separated in time. When used in this sense, the reference to a point in time should be clearly stated to avoid confusion given the varying other usages of the term.

# Latency, irreversibility, and persistence

*In the context of tipping transitions,* ***latency*** *refers to the time span between crossing the boundary between alternative regimes, and settling in the new regime.*

Due to inertia, there can indeed be a substantial delay between when a tipping transition begins (i.e., when the regime boundary is first crossed) and when its effects are fully realized. One of the most extreme examples here involves the polar ice sheets, where committed changes (already totaling multiple meters of sea-level rise) may play out over millennia after they are locked in because critical thresholds have been crossed. The presence of latency also implies that forcing crossing a critical (equilibrium) threshold does not necessarily imply an inevitable complete tipping transition; in some cases, the transition can be prevented if the forcing is reversed within a time span that is determined by the intrinsic time scales (and hence latency) of the system in question (Ritchie et al., 2021; Bochow et al., 2023).

*Because tipping transitions are self-sustaining transitions between alternative dynamical regimes, they can be* ***irreversible*** *in the sense that reversing the forcing or perturbation that triggered the* *transition may not reverse the transition even if the system is given sufficient time to reach equilibrium (so that all latency-related effects have already played out). The underlying dynamical phenomenon is called **hysteresis**.*

Bifurcation-induced tipping transitions are necessarily irreversible in this sense, and in this case, the irreversibility is associated with hysteresis and not with latency. Rate- and noise-induced transitions can but do not have to be irreversible, depending on the position of the system when the fast forcing or perturbation is applied, on the size of the alternative regime relative to the forcing change or perturbation, and on the time span between the forcing change or perturbation and their reversal.

The fact alone that something takes a long time (relative to human experience) does not render it irreversible and we emphasize that *irreversibility*, in the above sense of hysteresis, should not be confused with *latency*: The fact that a system does not reach the same state when a forcing is first applied over a finite time, and then reversed over a finite time, does not imply hysteresis and irreversibility in that sense. It could, rather, simply be caused by latency. To avoid confusion we note here that in the IPCC AR6 glossary a "perturbed state of a *dynamical system* is defined as irreversible on a given time scale if the recovery from this state due to natural processes takes substantially longer than the time scale of interest." (IPCC, 2021a) In our terminology, this would be latency and not irreversibility; we note that the IPCC AR6 glossary defines "irreversible on a given time scale" and recommend always making that specific reference to a time scale, or use the specific term "time-scale irreversibility" whenever confusion may arise. The difference between latency and irreversibility in the sense of hysteresis in tipping systems reflects the distinction between a system's stable equilibrium state and its transient response.

*Dynamical regimes as defined above are **persistent** in the sense that the system will remain in a given regime unless external factors drive it outside.*

We note here that persistence can have slightly different meanings both in the physical and the social science literature (Abrams et al., 2026). In physical systems, asymptotic persistence reflects multistability: up to a certain strength of perturbations, the system will remain in a given regime thanks to the acting attracting / stabilizing forces; only if perturbations exceed these forces will the system leave the regime. Metastability, in contrast, reflects persistence with respect to a given time scale. One could, in principle, generalize the above definition of dynamical regimes to also include regimes that would only be persistent over such a finite time scale (as done e.g. in (Börner & Dijkstra, 2026)). For example, in some ESM simulations the AMOC strongly reduces in response to hosing experiments and remains weak for centuries to millennia, before apparently spontaneously recovering to a strong circulation mode (Romanou et al., 2023). If this recovery is the result of some AMOC-internal mechanism and not triggered by an external perturbation, the weak AMOC state would not correspond to a dynamical regime as we define it above, and the simulated AMOC shifts would not be tipping transitions as defined above. However, when only requiring metastability and thus finite-time persistence *with respect to a certain time scale* to define dynamical regimes, these AMOC shifts would be tipping transitions. In social systems, concepts such as 'forcing,' 'perturbation,' 'multistability', or 'metastability' are challenging to define, but persistence within a regime can be associated with factors such as institutional or infrastructural lock-in (Abrams et al., 2026).

# Tipping transitions vs. abrupt shifts

Communication around tipping transitions can also be confusing because of ambiguity related to the term 'abrupt' that is often used to characterize tipping transitions. IPCC AR6 defines an abrupt change as “A change in the system that is substantially faster than the typical rate of the changes in its history,” which suggests a reference to time. The term ‘abrupt’ has also often been used with a reference to specific time scales. A typical reference that is often invoked is the “human time scale”. We note that tipping transitions are not generally fast compared to such human time scales, because they can have long latency due to long characteristic timescales. Describing tipping transitions as ‘abrupt’ could be confusing because many will assume that ‘abrupt’ is necessarily meant with respect to time and, potentially very misleadingly, expect an event that occurs rapidly with respect to human time scales; see also (Börner & Dijkstra, 2026; Macneill et al., 2026). Because of the subjectivity and relativity we generally recommend to not use the term abruptness in reference to a chosen time scale, but rather directly report that time scale; e.g., estimate how long a transition or shift takes.

By contrast, abruptness can be consistently defined with respect to forcing; a straightforward definition would be to call a shift or transition **abrupt** if the response amplifies nonlinearly with respect to forcing in a certain range of forcing level. In this sense, tipping transitions would indeed be abrupt with respect to forcing in quasi-equilibrium settings, where the forcing changes so slowly compared to characteristic system time scales that the system can equilibrate in response to each change in forcing (Fig. 1). We also refer to (Hansen et al., 2026) for a related attempt to define abruptness with respect to forcing, rather than time. However, the common perception of the term “abrupt” strongly leans toward a reference to time. Although it is less ambiguous, re-defining abruptness with respect to forcing would be in contrast with the colloquial meaning of the term, which may add further confusion. Moreover, defining abruptness as a nonlinear response to forcing would remain highly relative, since any change in response that is stronger than linear would be included. Because of the ambiguity and relativeness when defined with reference to time, and the remaining relativeness when defined with respect to forcing, we generally recommend to only use the term to describe specific events with clear reference to a time or forcing scale, and to avoid it as a general concept and for defining a general terminology.

There are strongly nonlinear shifts (which may even be considered abrupt with respect to decadal or shorter timescales) that are not tipping transitions because they do not occur between alternative regimes (Fig. 1a). Unlike tipping transitions, such nonlinear shifts are always reversible with a reversal of the drivers that triggered the transition. The loss of winter sea ice area in the central Arctic would be an example of this. While the retreat of the total hemispheric sea ice extent is linear over time when accounting for coastline geometry (Eisenman, 2010), ice coverage of a smaller region that has been completely ice covered in a pre-industrial climate can reduce nonlinearly with regard to forcing and time once the ocean temperature in the region under consideration reaches the freezing point. This threshold effect is particularly pronounced for Arctic winter sea ice because it has relatively homogeneous thickness compared to summer ice; hence a large area of sea ice can become open water in a short time (Bathiany et al., 2016). However, freezing and melting occurs at the same temperature threshold, which is why the transition is reversible if regional temperatures are reversed, even in a model with an extremely rapid winter sea ice loss (Li et al., 2013). The reason why the positive ice-albedo feedback does not lead to alternative states and hysteresis is related to the existence of the annual cycle and the fact that large negative feedbacks (most importantly, long-wave emission from the open ocean) stabilize the system (Tietsche et al., 2011; Wagner & Eisenman, 2015). We note that Arctic summer sea ice retreat would not be a nonlinear shift because when accounting for coastline geometry, it will respond linearly to regional temperature change (Intergovernmental Panel On Climate Change (Ipcc), 2023) (e.g., (Fox-Kemper et al., 2021).

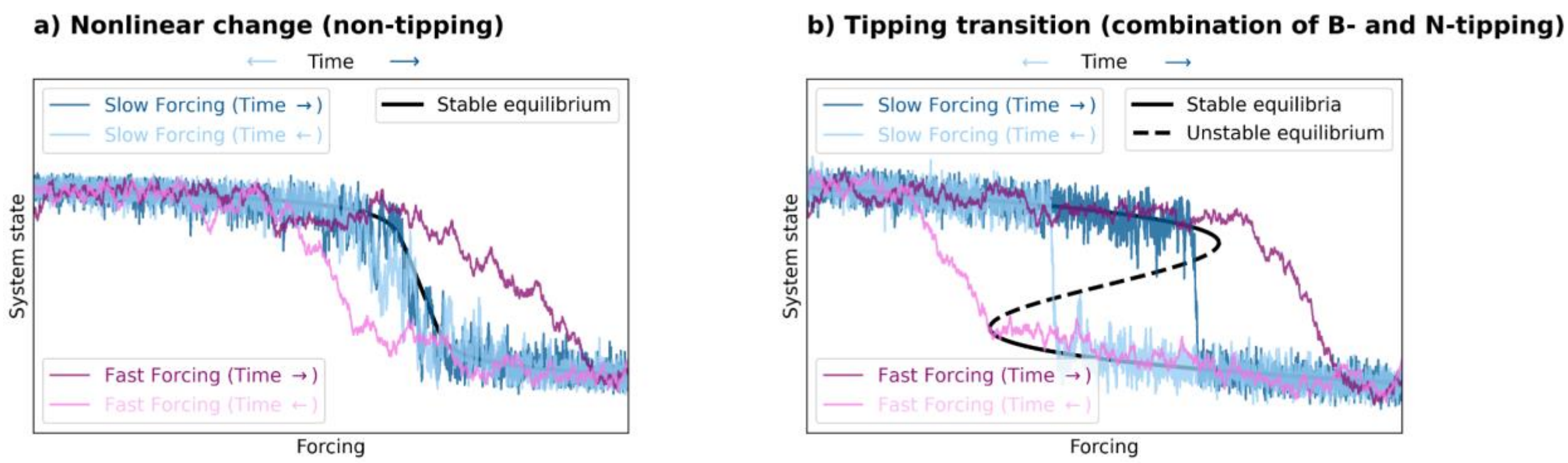

*Fig. 1: Nonlinear change vs. tipping transition. (a) Simulated example time series of a system undergoing a strongly nonlinear change for slow (blue) and fast (magenta) change in forcing from left to right (darker blue and magenta) and from right to left (lighter blue and magenta). The system is driven by additive white noise. The underlying attractor is depicted in solid black; there exists only one global regime associated with this attractor. The blue time series resulting from slow change in forcing show that this nonlinear shift is reversible in the sense that there is no hysteresis. The magenta time series resulting from fast forcing change show an apparent irreversibility, which is only due to latency, not dynamical irreversibility. This nonlinear shift would be called abrupt with respect to forcing according to the definition suggested here; however, the shift may or may not be considered abrupt with respect to time, depending on a reference time scale. (b) Simulated example time series of a tipping transition between alternative regimes. As in (a), slowly (rapidly) forced time series are shown in blue (magenta). The different attractors of the two alternative regimes are shown in solid black, while the regime boundary between the coexisting regimes is indicated by the black dashed line. In this case, even if the system is forced slowly and has time to approximately equilibrate with each change in forcing, the two transitions up and down do not occur for the same forcing, indicating hysteresis and thus irreversibility in a strict dynamical sense.*

# Interactions and tipping cascades

Tipping systems can also interact with one another. Consequently, a tipping transition in one system could lead to a tipping transition in another (Wunderling et al., 2021; Boers et al., 2025). A prominent example that has often been invoked in this context is that a tipping transition of the Greenland ice sheet and resulting freshwater release may trigger a tipping transition of the AMOC. However, it should be noted that examples of longer chains of transitions beyond three or four of the suggested large-scale tipping elements (sometimes referred to as tipping cascades) are not plausible to occur in the Earth system according to current understanding. One reason for this is that interactions can also have stabilizing effects; for example, a slowing AMOC would cause relative cooling in the Arctic and, e.g., lead to a stabilization of the Greenland ice sheet (Pöppelmeier & Stocker, 2025). A weakening AMOC has also been suggested to have stabilizing effects on eastern Amazon rainforests (Nian et al., 2023).

Paralleling the definition of 'compound extremes,' we can define compounding tipping transitions to occur when independent tipping transitions in separate systems lead to impacts that are more than linearly additive. A possible example would be an independent AMOC tipping and West Antarctic Ice Sheet (WAIS) tipping producing compound coastal impacts.

# Uncertainties

Tipping transitions are often associated with deep uncertainty, and therefore many fall into the category of high-impact-unknown-likelihood (HIUL) rather than high-impact-low-likelihood (HILL) events (Lempert et al., 2024). Here, we use the term HIUL to encompass all situations of high potential impact where likelihood is not well known enough to state conclusively that an event is of known low likelihood.  For example, an AMOC tipping transition is a key example of a HIUL event because of the large uncertainties in observational stability estimates (Ben-Yami et al., 2023, 2024) and persistent biases in state-of-the-art models (Liu et al., 2017; Mecking et al., 2017; Weijer et al., 2019). The distinction between HILL and HIUL events is important because some potential high-impact tipping transitions might in fact not be of low but rather of high likelihood to occur under plausible future forcing scenarios. For example, conditional on regional warming, the likelihood of a committed Greenland Ice Sheet meltdown is low and reasonably well constrained at low warming, deeply uncertain (i,.e., not well known) around the reported critical temperature range (Bochow et al., 2023; Robinson et al., 2012), and high at high warming. Thus, around the critical temperature range, committed Greenland Ice Sheet meltdown would be a HIUL event.

A different kind of uncertainty arises where the event itself is of known and even high likelihood, but it remains uncertain whether it qualifies as a tipping transition. This may have important implications, e.g. for the reversibility of the event. For example, warm-water coral reef dieoff is of neither poorly known nor low likelihood and it is often considered the canonical example of a tipping transition already underway. However, there is still uncertainty if warm-water coral reef dieoff really qualifies as a tipping transition or rather reflects a forced response to cumulative thermal stress; especially the question whether the alternative, algae-dominated state constitutes a dynamical regime in the sense defined above remains debated (Bruno et al., 2009; Dudgeon et al., 2010; Hughes et al., 2017; Żychaluk et al., 2012) (Mumby et al., 2007; Muthukrishnan et al., 2016).

Social systems are characterized by uncertainties that are not only deep but reflexive, in that their structure and magnitude depend on people's actions, which are shaped by beliefs, values, and norms, and on the way each actor's conduct is conditioned by their anticipation of others. Social scientists disagree about how far a logic of thresholds in social systems can be carried: positivist traditions may treat social thresholds as quasi-deterministic and analogous to thresholds in system dynamics, while interpretive and agency-centred perspectives take human systems to be more reflexive, intentional, and open and therefore resist importing a deterministic threshold logic. When using 'tipping' terminology to describe social dynamics one should therefore at least check for metaphorical consistency with the core traits identified here, and be clear that such terminology — including the identification of system and regime boundaries, elements, and agency it requires — may not be normatively neutral.

It is also important to recognize that, while in idealized models of tipping systems it may be possible to define regimes and their boundaries objectively, in real-world applications, defining regime boundaries involves a high degree of social construction. While this is fairly clear in social systems, it is true even in many physical systems, as in the challenge of defining what constitutes AMOC ‘collapse’, or when AMOC has initiated a tipping transition, or even in drawing boundaries between ice sheets and other elements of the Earth system (e.g., Sergienko et al., 2026) .

# Conclusion

In recent years, in both the scientific literature and public debate regarding climate change, the term ‘tipping point’ has seen greatly increasing yet inconsistent usage. To address this inconsistency, we have attempted to provide self-contained definitions of ‘tipping transitions’ and related terms that are consistent with the underlying mathematical terminology (bifurcations, attractors, basins of attraction, etc.), but more accessible to a wider audience and applicable across natural and social sciences. Specifically, we define a ‘tipping transition’ as a *self-sustaining transition between alternative dynamical regimes.*

Because of its contested definitions – a contest that some of the authors of this essay regard as irresolvable because of the term’s now well established roots in popular culture – we argue that the specific notion of a ‘tipping point’ remains ambiguous and recommend using other, more suitable and precise terms as laid out in this essay. Although the term can be made more precise with reference to the specific point in time when a tipping transition is triggered, we suggest it be more generally regarded as an informal, metaphorical term, sitting at the boundary of several different disciplinary and cultural perspectives (Kopp et al., 2025). While the value of the term for general communication is debated (Abrams et al., 2026; Kopp et al., 2025; Smith et al., 2025), other technical terms, such as multiple equilibria, bifurcations (Ghil, 2001), critical transitions (Kuehn, 2011) and their generalizations (Ghil & Lucarini, 2020; Lucarini & Chekroun, 2024), as well as the terminology outlined in this essay, should be better suited for precise technical communication and scientific writing.

We also argue that the notion of ‘high-impact-low-likelihood (HILL) events’ can be misleading when used to describe potential tipping transitions, since such transitions are typically of poorly known rather than of low likelihood; a more precise characterization would thus be high-impact-unknown-likelihood (HIUL) events (Kopp et al., 2025; Lempert et al., 2024). Several previous studies have defined tipping transitions based on the notions of abruptness and irreversibility. However, we have shown that these terms are not straightforward to define themselves, especially if understood with reference to time rather than to forcing. Even if made precise with reference to forcing as indicated above, abruptness and irreversibility are not generally necessary or sufficient conditions for tipping transitions (Börner & Dijkstra, 2026). We have therefore chosen to define tipping transitions independently of these terms, while explaining the relationship and acknowledging that the potential irreversibility and abruptness (with respect to a certain time scale) of tipping transitions in the Earth system are important traits that can be highly relevant in terms of possible tipping transitions in the context of anthropogenic climate change.

## Acknowledgements

This is ClimTip contribution \#; the ClimTip project has received funding from the European Union's Horizon Europe research and innovation programme under Grant 101137601: Funded by the European Union. NB acknowledges further funding by the Volkswagen Foundation, the European Space Agency Climate Change Initiative (ESA-CCI) Tipping Elements SIRENE project (contract no. 4000146954/24/I-LR), and the Past to Future (P2F) project, which has received funding from the European Union's Horizon Europe research and innovation programme under grant agreement No. 101184070. REK was supported by by the National Science Foundation as part of the Megalopolitan Coastal Transformation Hub (MACH) under award ICER-2103754 and by the National Aeronautics and Space Administration (NASA) Sea-level Change Team. TLF was supported by the Horizon Europe project TipESM "Exploring Tipping Points and Their Impacts Using Earth System Models" funded by the European Union (grant agreement no. 101137673. DOI: 10.3030/101137673, TipESM contribution no. XX). VL and PA acknowledge the partial support provided by the Horizon Europe Projects Past2Future (Grant No. 101184070) and ClimTIP (Grant No. 100018693), by the ARIA SCOP-PR01-P003 - Advancing Tipping Point Early Warning AdvanTip project, and by the European Space Agency Project PREDICT (Contract 4000146344/24/I-LR). VL also acknowledges the partial support received as Simons Fellow of the I. Newton Institute (FCPW01 programme). SJ acknowledges Horizon Europe Project FUTURA (101295156) for her time.

## References

Abrams, J. F., Pereira, L. M., Spaiser, V., Wunderling, N., Villasante, S., Tàbara, J. D., Biggs, R., Armstrong McKay, D. I., Zimm, C., Milkoreit, M., Constantino, S. M., Debnath, R., Dakos, V., Ritchie, P. D. L., Smith, S. R., Rocha, J. C., Kéfi, S., Buxton, J. E., Boulton, C. A., … Lenton, T. M. (2026). Integrating tipping point concepts across diverse systems. *Communications Sustainability*, *1*(1), 68. https://doi.org/10.1038/s44458-026-00063-5

Alkhayuon, H., Ashwin, P., Jackson, L. C., Quinn, C., & Wood, R. A. (2019). Basin bifurcations, oscillatory instability and rate-induced thresholds for Atlantic meridional overturning circulation in a global oceanic box model. *Proceedings of the Royal Society A: Mathematical, Physical and Engineering Sciences*, *475*(2225), 20190051. https://doi.org/10.1098/rspa.2019.0051

Armstrong McKay, D. I., Staal, A., Abrams, J. F., Winkelmann, R., Sakschewski, B., Loriani, S., Fetzer, I., Cornell, S. E., Rockström, J., & Lenton, T. M. (2022). Exceeding 1.5°C global warming could trigger multiple climate tipping points. *Science*, *377*(6611), eabn7950. https://doi.org/10.1126/science.abn7950

Ashwin, P., Feudel, U., Ghil, M., Lehnertz, K., Ortega, J.-P., & Rasmussen, M. (2026). Introduction to Focus Issue: Nonautonomous dynamical systems: Theory, methods, and applications. *Chaos: An Interdisciplinary Journal of Nonlinear Science*, *36*(4), 040401. https://doi.org/10.1063/5.0324359

Ashwin, P., Wieczorek, S., Vitolo, R., & Cox, P. (2012). Tipping points in open systems: Bifurcation, noise-induced and rate-dependent examples in the climate system. *Philosophical Transactions of the Royal Society A: Mathematical, Physical and Engineering Sciences*, *370*(1962), 1166–1184. https://doi.org/10.1098/rsta.2011.0306

Bathiany, S., Notz, D., Mauritsen, T., Raedel, G., & Brovkin, V. (2016). On the Potential for Abrupt Arctic Winter Sea Ice Loss. *Journal of Climate*, *29*(7), 2703–2719. https://doi.org/10.1175/JCLI-D-15-0466.1

Ben-Yami, M., Morr, A., Bathiany, S., & Boers, N. (2024). Uncertainties too large to predict tipping times of major Earth system components from historical data. *Science Advances*, *10*(31), eadl4841. https://doi.org/10.1126/sciadv.adl4841

Ben-Yami, M., Skiba, V., Bathiany, S., & Boers, N. (2023). Uncertainties in critical slowing down indicators of observation-based fingerprints of the Atlantic Overturning Circulation. *Nature Communications*, *14*(1), 8344. https://doi.org/10.1038/s41467-023-44046-9

Bochow, N., & Boers, N. (2023). The South American monsoon approaches a critical transition in response to deforestation. *Science Advances*, *9*(40), eadd9973. https://doi.org/10.1126/sciadv.add9973

Bochow, N., Poltronieri, A., Robinson, A., Montoya, M., Rypdal, M., & Boers, N. (2023). Overshooting the critical threshold for the Greenland ice sheet. *Nature*, *622*(7983), 528–536. https://doi.org/10.1038/s41586-023-06503-9

Boers, N., Ghil, M., & Stocker, T. F. (2022). Theoretical and paleoclimatic evidence for abrupt transitions in the Earth system. *Environmental Research Letters*, *17*(9), 093006. https://doi.org/10.1088/1748-9326/ac8944

Boers, N., Liu, T., Bathiany, S., Ben-Yami, M., Blaschke, L. L., Bochow, N., Boulton, C. A., Lenton, T. M., Morr, A., Nian, D., Rypdal, M., & Smith, T. (2025). Destabilization of Earth system tipping elements. *Nature Geoscience*, *18*(10), 949–960. https://doi.org/10.1038/s41561-025-01787-0

Börner, R., & Dijkstra, H. A. (2026). *To tip or not to tip*. Climate change/Human/Earth system interactions/Mathematical geosciences and physical models. https://doi.org/10.5194/egusphere-2026-3507

Bruno, J. F., Sweatman, H., Precht, W. F., Selig, E. R., & Schutte, V. G. W. (2009). Assessing evidence of phase shifts from coral to macroalgal dominance on coral reefs. *Ecology*, *90*(6), 1478–1484. https://doi.org/10.1890/08-1781.1

Budyko, M. I. (1969). The effect of solar radiation variations on the climate of the Earth. *Tellus*, *21*(5), 611–619. https://doi.org/10.1111/j.2153-3490.1969.tb00466.x

Buhr, L., Lenzi, D. S., Pols, A. J. K., Brunner, C. E., Fischer, A., Staal, A., Hofbauer, B. P., & Bovenkerk, B. (2025). The concepts of irreversibility and reversibility in research on anthropogenic environmental changes. *PNAS Nexus*, *4*(1), pgae577. https://doi.org/10.1093/pnasnexus/pgae577

Chekroun, M. D., Simonnet, E., & Ghil, M. (2011). Stochastic climate dynamics: Random attractors and time-dependent invariant measures. *Physica D: Nonlinear Phenomena*, *240*(21), 1685–1700. https://doi.org/10.1016/j.physd.2011.06.005

Cini, M., Zappa, G., Ragone, F., & Corti, S. (2024). Simulating AMOC tipping driven by internal climate variability with a rare event algorithm. *Npj Climate and Atmospheric Science*, *7*(1), 31. https://doi.org/10.1038/s41612-024-00568-7

Crisan, D., Galatolo, S., Ghil, M., Pierini, S., Sciamarella, D., & Tél, T. (2026). Introduction to the Focus Issue: Nonautonomous dynamics in the climate sciences. *Chaos: An*

*Interdisciplinary Journal of Nonlinear Science*, *36*(4), 040403. https://doi.org/10.1063/5.0324361
Dansgaard, W. (1987). Ice core evidence of abrupt climatic changes. In *Abrupt climatic change. Proc. 1985* (pp. 223–233).
Dijkstra, H. A., & Ghil, M. (2005). Low-frequency variability of the large-scale ocean circulation: A dynamical systems approach. *Reviews of Geophysics*, *43*(3), 1–38. https://doi.org/10.1029/2002RG000122
Ditlevsen, P. D., Andersen, K. K., & Svensson, A. (2007). The DO-climate events are probably noise induced: Statistical investigation of the claimed 1470 years cycle. *Climate of the Past*, *3*, 129–134. https://doi.org/10.5194/cpd-2-1277-2006
Dudgeon, S. R., Aronson, R. B., Bruno, J. F., & Precht, W. F. (2010). Phase shifts and stable states on coral reefs. *Marine Ecology Progress Series*, *413*, 201–216. https://doi.org/10.3354/meps08751
Eisenman, I. (2010). Geographic muting of changes in the Arctic sea ice cover. *Geophysical Research Letters*, *37*(16), 2010GL043741. https://doi.org/10.1029/2010GL043741
Flores, B. M., Montoya, E., Sakschewski, B., Nascimento, N., Staal, A., Betts, R. A., Levis, C., Lapola, D. M., Esquível-Muelbert, A., Jakovac, C., Nobre, C. A., Oliveira, R. S., Borma, L. S., Nian, D., Boers, N., Hecht, S. B., ter Steege, H., Arieira, J., Lucas, I. L., … Hirota, M. (2024). Critical transitions in the Amazon forest system. *Nature*, *626*(7999), 555–564. https://doi.org/10.1038/s41586-023-06970-0
Fox-Kemper, B., H.T. Hewitt, C. Xiao, G. Aðalgeirsdóttir, S.S. Drijfhout, T.L. Edwards, N.R. Golledge, M. Hemer, R.E. Kopp, G. Krinner, A. Mix, D. Notz, S. Nowicki, I.S. Nurhati, L. Ruiz, J.-B. Sallée, A.B.A. Slangen, and Y. Yu. (n.d.). Ocean, Cryosphere and Sea Level Change. In Masson-Delmotte, V., P., Zhai, A. Pirani, S.L. Connors, C. Péan, S. Berger, N. Caud, Y. Chen, L. Goldfarb, M.I. Gomis, M. Huang, K. Leitzell, E., Lonnoy, J.B.R. Matthews, T.K. Maycock, T. Waterfield, O. Yelekçi, R. Yu, and B. Zhou (Ed.), *Limate Change 2021: The Physical Science Basis. Contribution of Working Group I to the Sixth Assessment Report of the Intergovernmental Panel on Climate Change* (pp. 1211–1362). Cambridge University Press, Cambridge. https://doi.org/10.1017/9781009157896.011
Ghil, M. (1976). Climate Stability for a Sellers-Type Model. *Journal of the Atmospheric Sciences*, *33*(1), 3–20. https://doi.org/10.1175/1520-0469(1976)033%3C0003:CSFAST%3E2.0.CO;2
Ghil, M. (2001). Hilbert problems for the geosciences in the 21st century. *Nonlinear Processes in Geophysics*, *8*(4/5), 211–211. https://doi.org/10.5194/npg-8-211-2001
Ghil, M., & Lucarini, V. (2020). The physics of climate variability and climate change. *Reviews of Modern Physics*, *92*(3), 035002. https://doi.org/10.1103/RevModPhys.92.035002
Gladwell, M. (2000). *The Tipping Point: How Little Things Can Make a Big Difference*. Abacus.
Hansen, G., van der Wel, N., Abbott, P. M., Naish, T., & Vautard, R. (2026). *Proceedings from WCRP-IPCC Co-Sponsored workshop on Earth System High Impact Events, Tipping Points and Their Consequences* (2026th ed.). World Meteorological Organization. https://doi.org/10.59327/WMO/WCRP/IPCC/HIETP
Hughes, T. P., Barnes, M. L., Bellwood, D. R., Cinner, J. E., Cumming, G. S., Jackson, J. B. C., Kleypas, J., van de Leemput, I. A., Lough, J. M., Morrison, T. H., Palumbi, S. R., van Nes, E. H., & Scheffer, M. (2017). Coral reefs in the Anthropocene. *Nature*, *546*(7656), 82–90. https://doi.org/10.1038/nature22901
Intergovernmental Panel On Climate Change (Ipcc). (2023). *Climate Change 2021 – The Physical Science Basis: Working Group I Contribution to the Sixth Assessment Report of the Intergovernmental Panel on Climate Change* (1st ed.). Cambridge University Press. https://doi.org/10.1017/9781009157896

IPCC. (2021a). *Climate Change 2021: The Physical Science Basis. Contribution of Working Group I to the Sixth Assessment Report of the Intergovernmental Panel on Climate Change* (S. L. Masson-Delmotte, V., P. Zhai, A. Pirani, J. B. R. Connors, C. Péan, S. Berger, N. Caud, Y. Chen, L. Goldfarb, M.I. Gomis, M. Huang, K. Leitzell, E. Lonnoy, Matthews, T.K. Maycock, T. Waterfield, O. Yelekçi, R. Yu, & B. Zhou (eds.), Eds.). Cambridge University Press.

IPCC. (2021b). Summary for Policymakers. In V. Masson-Delmotte, P. Zhai, A. Pirani, S. L. Connors, C. Péan, S. Berger, N. Caud, Y. Chen, L. Goldfarb, M. I. Gomis, M. Huang, K. Leitzell, E. Lonnoy, J. B. R. Matthews, T. K. Maycock, T. Waterfield, O. Yelekçi, R. Yu, & B. Zhou (Eds.), *Climate change 2021: The physical science basis* (pp. 3–32). Cambridge University Press. https://doi.org/10.1017/9781009157896.001

Jeppesen, E., Søndergaard, M., Jensen, J. P., Havens, K. E., Anneville, O., Carvalho, L., Coveney, M. F., Deneke, R., Dokulil, M. T., Foy, B., Gerdeaux, D., Hampton, S. E., Hilt, S., Kangur, K., Köhler, J., Lammens, E. H. h. r., Lauridsen, T. L., Manca, M., Miracle, M. R., … Winder, M. (2005). Lake responses to reduced nutrient loading – an analysis of contemporary long-term data from 35 case studies. *Freshwater Biology*, *50*(10), 1747–1771. https://doi.org/10.1111/j.1365-2427.2005.01415.x

Kopp, R. E., Gilmore, E. A., Shwom, R. L., Adams, H., Adler, C., Oppenheimer, M., Patwardhan, A., Russill, C., Schmidt, D. N., & York, R. (2025). 'Tipping points' confuse and can distract from urgent climate action. *Nature Climate Change*, *15*(1), 29–36. https://doi.org/10.1038/s41558-024-02196-8

Kopp, R. E., Oppenheimer, M., O'Reilly, J. L., Drijfhout, S. S., Edwards, T., Fox-Kemper, B., Garner, G. G., Golledge, N. R., Hermans, T., Hewitt, H. T., Horton, B. P., Krinner, G., Notz, D., Nowicki, S., Palmer, M. D., & Slangen, A. B. A. (2023). Communicating future sea-level rise uncertainty and ambiguity to assessment users. *Nature Climate Change*, *13*, 648–660. https://doi.org/10.1038/s41558-023-01691-8

Kopp, R. E., Shwom, R., Wagner, G., & Yuan, J. (2016). Tipping elements and climate-economic shocks: Pathways toward integrated assessment. *Earth's Future*, *4*, 346–372. https://doi.org/10.1002/2016EF000362.

Kramers, H. A. (1940). Brownian motion in a field of force and the diffusion model of chemical reactions. *Physica*, *7*(4), 284–304. https://doi.org/10.1016/S0031-8914(40)90098-2

Kuehn, C. (2011). A mathematical framework for critical transitions: Bifurcations, fast–slow systems and stochastic dynamics. *Physica D: Nonlinear Phenomena*, *240*(12), 1020–1035. https://doi.org/10.1016/j.physd.2011.02.012

Lempert, R. J., Lawrence, J., Kopp, R. E., Haasnoot, M., Reisinger, A., Grubb, M., & Pasqualino, R. (2024). The use of decision making under deep uncertainty in the IPCC. *Frontiers in Climate*, *6*. https://doi.org/10.3389/fclim.2024.1380054

Lenton, T. M., Held, H., Kriegler, E., Hall, J. W., Lucht, W., Rahmstorf, S., & Schellnhuber, H. J. (2008). Tipping elements in the Earth's climate system. *Proceedings of the National Academy of Sciences*, *105*(6), 1786–1793. https://doi.org/10.1073/pnas.0705414105

Li, C., Notz, D., Tietsche, S., & Marotzke, J. (2013). The Transient versus the Equilibrium Response of Sea Ice to Global Warming. *Journal of Climate*, *26*(15), 5624–5636. https://doi.org/10.1175/JCLI-D-12-00492.1

Liu, W., Xie, S.-P., Liu, Z., & Zhu, J. (2017). Overlooked possibility of a collapsed Atlantic Meridional Overturning Circulation in warming climate. *Science Advances*, *3*(1), e1601666. https://doi.org/10.1126/sciadv.1601666

Lucarini, V., Calmanti, S., & Artale, V. (2007). Experimental mathematics: Dependence of the stability properties of a two-dimensional model of the Atlantic ocean circulation on the boundary conditions. *Russian Journal of Mathematical Physics*, *14*(2), 224–231. https://doi.org/10.1134/S1061920807020124

Lucarini, V., & Chekroun, M. D. (2024). Detecting and Attributing Change in Climate and Complex Systems: Foundations, Green's Functions, and Nonlinear Fingerprints. *Physical Review Letters*, *133*(24), 244201. https://doi.org/10.1103/PhysRevLett.133.244201

Macneill, K., Martin, L., Wood, R. A., & Sutton, R. (2026). An Up-HILL Battle: Towards clearer terminology for High-Impact Climate Possibilities, including tipping points. *EGUsphere*, 1–37. https://doi.org/10.5194/egusphere-2026-3767

Mecking, J. V., Drijfhout, S. S., Jackson, L. C., & Andrews, M. B. (2017). The effect of model bias on Atlantic freshwater transport and implications for AMOC bi-stability. *Tellus, Series A: Dynamic Meteorology and Oceanography*, *69*(1), 1–15. https://doi.org/10.1080/16000870.2017.1299910

Milkoreit, M., Hodbod, J., Baggio, J., Benessaiah, K., Calderón-Contreras, R., Donges, J. F., Mathias, J.-D., Rocha, J. C., Schoon, M., & Werners, S. E. (2018a). Defining tipping points for social-ecological systems scholarship—An interdisciplinary literature review. *Environmental Research Letters*, *13*(3), 033005. https://doi.org/10.1088/1748-9326/aaaa75

Milkoreit, M., Hodbod, J., Baggio, J., Benessaiah, K., Calderón-Contreras, R., Donges, J. F., Mathias, J.-D., Rocha, J. C., Schoon, M., & Werners, S. E. (2018b). Defining tipping points for social-ecological systems scholarship—An interdisciplinary literature review. *Environmental Research Letters*, *13*(3), 033005. https://doi.org/10.1088/1748-9326/aaaa75

Mumby, P. J., Hastings, A., & Edwards, H. J. (2007). Thresholds and the resilience of Caribbean coral reefs. *Nature*, *450*(7166), 98–101. https://doi.org/10.1038/nature06252

Muthukrishnan, R., Lloyd-Smith, J. O., & Fong, P. (2016). Mechanisms of resilience: Empirically quantified positive feedbacks produce alternate stable states dynamics in a model of a tropical reef. *Journal of Ecology*, *104*(6), 1662–1672. https://doi.org/10.1111/1365-2745.12631

Nian, D., Bathiany, S., Ben-Yami, M., Blaschke, L. L., Hirota, M., Rodrigues, R. R., & Boers, N. (2023). A potential collapse of the Atlantic Meridional Overturning Circulation may stabilise eastern Amazonian rainforests. *Communications Earth & Environment*, *4*(1), 470. https://doi.org/10.1038/s43247-023-01123-7

North, G. R. (1975). Theory of Energy-Balance Climate Models. *Journal of the Atmospheric Sciences*, *32*(11), 2033–2043. https://doi.org/10.1175/1520-0469(1975)032%3C2033:TOEBCM%3E2.0.CO;2

Peixoto, J. P., & Oort, A. H. (1993). *Physics of climate* (3. print). American Institute of Physics.

Peltier, W. R., & Vettoretti, G. (2014). Dansgaard-Oeschger oscillations predicted in a comprehensive model of glacial climate: A "kicked" salt oscillator in the Atlantic. *Geophysical Research Letters*, *41*(20), 7306–7313. https://doi.org/10.1002/2014GL061413

Pöppelmeier, F., & Stocker, T. F. (2025). Mutual stabilization of AMOC and GrIS due to different transient response to warming. *Environmental Research Letters*, *20*(9), 094015. https://doi.org/10.1088/1748-9326/adf45a

Ritchie, P. D. L., Alkhayuon, H., Cox, P. M., & Wieczorek, S. (2023). Rate-induced tipping in natural and human systems. *Earth System Dynamics*, *14*(3), 669–683. https://doi.org/10.5194/esd-14-669-2023

Ritchie, P. D. L., Clarke, J. J., Cox, P. M., & Huntingford, C. (2021). Overshooting tipping point thresholds in a changing climate. *Nature*, *592*(7855), 517–523. https://doi.org/10.1038/s41586-021-03263-2

Robinson, A., Calov, R., & Ganopolski, A. (2012). Multistability and critical thresholds of the Greenland ice sheet. *Nature Climate Change*, *2*(6), 429–432. https://doi.org/10.1038/nclimate1449
Romanou, A., Hegerl, G. C., Seneviratne, S. I., Abis, B., Bastos, A., Conversi, A., Landolfi, A., Kim, H., Lerner, P. E., Mekus, J., Otto-Bliesner, B. L., Pausata, F. S. R., Pinto, I., & Suarez-Guiterrez, L. (2025). Extreme Events Contributing to Tipping Elements and Tipping Points. *Surveys in Geophysics*, *46*(2), 375–420. https://doi.org/10.1007/s10712-024-09863-7
Romanou, A., Rind, D., Jonas, J., Miller, R., Kelley, M., Russell, G., Orbe, C., Nazarenko, L., Latto, R., & Schmidt, G. A. (2023). Stochastic Bifurcation of the North Atlantic Circulation under a Midrange Future Climate Scenario with the NASA-GISS ModelE. *Journal of Climate*, *36*(18), 6141–6161. https://doi.org/10.1175/JCLI-D-22-0536.1
Russill, C. (2015). Climate change tipping points: Origins, precursors, and debates. *WIREs Climate Change*, *6*(4), 427–434. https://doi.org/10.1002/wcc.344
Russill, C., & Nyssa, Z. (2009). The tipping point trend in climate change communication. *Global Environmental Change*, *19*(3), 336–344. https://doi.org/10.1016/j.gloenvcha.2009.04.001
Scheffer, M. (2009). *Critical Transitions in Nature and Society*. Princeton University Press. https://doi.org/https://doi.org/10.2307/j.ctv173f1g1
Scheffer, M., Carpenter, S., Foley, J. A., Folke, C., & Walker, B. (2001). Catastrophic shifts in ecosystems. *Nature*, *413*(6856), 591–596. https://doi.org/10.1038/35098000
Scheffer, M., Hosper, S. H., Meijer, M.-L., Moss, B., & Jeppesen, E. (1993). Alternative equilibria in shallow lakes. *Trends in Ecology & Evolution*, *8*(8), 275–279. https://doi.org/10.1016/0169-5347(93)90254-M
Schelling, T. C. (1971). Dynamic models of segregation. *Journal of Mathematical S*, *1*(2), 143–186. http://www.tandfonline.com/doi/abs/10.1080/0022250X.1971.9989794
Sellers, W. D. (1969). A Global Climatic Model Based on the Energy Balance of the Earth-Atmosphere System. *Journal of Applied Meteorology and Climatology*, *8*(3), 392–400. https://doi.org/10.1175/1520-0450(1969)008%3C0392:AGCMBO%3E2.0.CO;2
Smith, S. R., Milkoreit, M., Geels, F. W., & Lenton, T. M. (2025). Advancing science, policy and action in tipping points research. *Nature Climate Change*, *15*(6), 576–577. https://doi.org/10.1038/s41558-025-02335-9
Stocker, T. F., & Schmittner, A. (1997). Influence of CO2 emission rates on the stability of the thermohaline circulation. *Nature*, *388*(6645), 862–865. https://doi.org/10.1038/42224
Stommel, H. (1961). Thermohaline Convection with Two Stable Regimes of Flow. *Tellus*, *13*(2), 224–230. https://doi.org/10.1111/j.2153-3490.1961.tb00079.x
Tietsche, S., Notz, D., Jungclaus, J. H., & Marotzke, J. (2011). Recovery mechanisms of Arctic summer sea ice: RECOVERY MECHANISMS OF ARCTIC SUMMER SEA ICE. *Geophysical Research Letters*, *38*(2), n/a-n/a. https://doi.org/10.1029/2010GL045698
Van Nes, E. H., Arani, B. M. S., Staal, A., Van Der Bolt, B., Flores, B. M., Bathiany, S., & Scheffer, M. (2016). What Do You Mean, 'Tipping Point'? *Trends in Ecology & Evolution*, *31*(12), 902–904. https://doi.org/10.1016/j.tree.2016.09.011
Van Westen, R. M., Börner, R., & Dijkstra, H. A. (2026). Failure to track a stable AMOC state under rapid climate change. *Nature Climate Change*. https://doi.org/10.1038/s41558-026-02730-w
Vettoretti, G., Ditlevsen, P., Jochum, M., & Rasmussen, S. O. (2022). Atmospheric CO2 control of spontaneous millennial-scale ice age climate oscillations. *Nature Geoscience*, *15*(4), 300–306. https://doi.org/10.1038/s41561-022-00920-7
Wagner, T. J. W., & Eisenman, I. (2015). How Climate Model Complexity Influences Sea Ice Stability. *Journal of Climate*, *28*(10), 3998–4014. https://doi.org/10.1175/JCLI-D-14-00654.1

Weijer, W., Cheng, W., Drijfhout, S. S., Fedorov, A. V., Hu, A., Jackson, L. C., Liu, W., McDonagh, E. L., Mecking, J. V., & Zhang, J. (2019). Stability of the Atlantic Meridional Overturning Circulation: A Review and Synthesis. *Journal of Geophysical Research: Oceans*, *124*(8), 5336–5375. https://doi.org/10.1029/2019JC015083

Wieczorek, S., Xie, C., & Ashwin, P. (2023). Rate-induced tipping: Thresholds, edge states and connecting orbits. *Nonlinearity*, *36*(6), 3238–3293. https://doi.org/10.1088/1361-6544/accb37

Wood, R. A., Crucifix, M., Lenton, T. M., Mach, K. J., Moore, C., New, M., Sharpe, S., Stocker, T. F., & Sutton, R. T. (2023). A Climate Science Toolkit for High Impact-Low Likelihood Climate Risks. *Earth's Future*, *11*(4), e2022EF003369. https://doi.org/10.1029/2022EF003369

Wunderling, N., Donges, J. F., Kurths, J., & Winkelmann, R. (2021). Interacting tipping elements increase risk of climate domino effects under global warming. *Earth System Dynamics*, *12*(2), 601–619. https://doi.org/10.5194/esd-12-601-2021

Zhang, W., Huang, Y., Bathiany, S., Shin, Y., Ben-Yami, M., Zhou, S., & Boers, N. (2026). Probabilistic anticipation of AMOC transitions under critical forcing magnitudes and rates via deep learning. *Chaos, Solitons & Fractals*, *209*, 118374. https://doi.org/10.1016/j.chaos.2026.118374

Żychaluk, K., Bruno, J. F., Clancy, D., McClanahan, T. R., & Spencer, M. (2012). Data-driven models for regional coral-reef dynamics. *Ecology Letters*, *15*(2), 151–158. https://doi.org/10.1111/j.1461-0248.2011.01720.x

Xiao C., Zhang T., Che T., et al. Do abrupt cryosphere events in High Mountain Asia indicate earlier tipping point than expected? Advances in Climate Change Research, 2023, 14(6):873-883. DOI:10.1016/j.accre.2023.11.006.